\documentclass[pra,aps,twocolumn]{revtex4}

\usepackage{latexsym}
\usepackage{amsmath}
\usepackage{bm}
\usepackage{bm,color}

\usepackage{graphicx}

\newcommand{\be}{\begin{eqnarray}}
\newcommand{\ee}{\end{eqnarray}}

\renewcommand{\theequation}{\arabic{equation}}

\begin{document}
\title{Vortex dynamics  in lattice Bose gases in
a synthesized magnetic field with a random noise
and a dissipation: Study by the stochastic Gross-Pitaevskii equation}

\date{\today}
\author{Ryosuke Kato} 
\author{Yoshihito Kuno} 
\author{Ikuo Ichinose}
\affiliation{%
Department of Applied Physics, Nagoya Institute of Technology,
Nagoya, 466-8555 Japan}
\begin{abstract}
In this paper, we investigate vortex dynamics in a two-dimensional Bose-Hubbard model 
coupled with a weak artificial magnetic field, a random white noise and a dissipation.
Origin of the noise and dissipation is considered as thermal fluctuations of
atoms that do not participate the Bose-Einstein condensation (BEC).
Solving a stochastic Gross-Pitaevskii equation to this system, 
we show that the interplay of the magnetic field and the white noise generates
vortices in the bulk of the BEC and stable steady states of vortices form 
after a transition period.
We calculate the incompressible part of the kinetic-energy spectrum of the BEC.
In the transition period,
a Kolmogorov $k^{-5/3}$ spectrum appears in the infrared regime with 
the wave number $k$, $k<\zeta^{-1}$, where $\zeta$ is the healing length, whereas
in the ultraviolet region, $k>\zeta^{-1}$, the spectrum behaves as $k^{-3}$.
On the other hand in the steady states, another scaling low appears.
We find a relationship between the above mentioned kinetic-energy spectra and 
the velocity of vortices.
By an inverse cascade, the large velocity of a few created vortices develops 
the Kolmogorov $k^{-5/3}$ spectrum.
\end{abstract}

\maketitle
\section{Introduction}

Study on the cold atomic system is one of the most active fields in physics 
nowadays \cite{Pethick,Lewenstein}.
In particular, the vortex dynamics in atomic Bose-Einstein condensation (BEC) is
investigated intensively by experiments \cite{Walmsley,Maurer}, and also
there are various theoretical studies on it \cite{Tsubota_rev,Vinen,Tsubota}. 
Among various phenomena including turbulence, vortices play a very important role 
in classical and quantum fluid dynamics.
Recently, vortices and turbulence in BECs of cold atoms 
are studied experimentally and theoretically and interesting results have been
obtained.
The cold atomic gases are highly controllable in experiments
and the creation and observation of turbulence cascade are easier than other 
quantum systems.
In particular, certain efficient methods to inject energy into BEC systems have been
proposed and used in experiments \cite{Bradley2}.
 
Theoretical and experimental investigation on quantum turbulence in cold
atomic gases performed so far is summarized as follows. 
In the three-dimensional (3D) case, a random vortex tangle is an essential 
feature of quantum turbulence and the Richardoson cascade is observed 
as typical dynamics of turbulence \cite{Tsubota_rev,Kobayashi1,Kobayashi2}.
In particular in a compressible quantum fluid, vortex core plays an important
role and the healing length $\zeta$ gives some specific length scale characterizing 
superfluid (SF).
Inertial ranges with the Kolmogorov $k^{-5/3}$ spectrum exist
in the infrared regime $k<\zeta^{-1}$.
In the 3D case, this behavior is explained by the direct energy cascade
from large scale to small scale through the Richardson mechanism.
On the other hand for the ultraviolet regime $k>\zeta^{-1}$, an acoustic
excitation in vortex cores gives $k^{-3}$ behavior \cite{Bradley1}.
In the 2D case, similar spectrum structure is observed, but an important 
difference from the 3D case is the existence of the {\em inverse cascade
phenomenon}.
That is, energy injected near $k\sim \zeta^{-1}$ moves toward 
the large scale region $k\ll \zeta^{-1}$ and influences configurations of vortices. 
Existence of the inertial range with the Kolmogorov $k^{-5/3}$ spectrum attributes 
to the vortex distribution in 2D BECs \cite{Bradley2,Skaugen}.

In this paper, we study the behavior of quantized vortices and their dynamics
in 2D cold atomic systems.
In particular, we inject energies into the BEC system in {\em small length 
scale by a random noise} and also apply a weak uniform magnetic field.
It is usually expected that the random noise at scales smaller than 
the healing length $\zeta$ only effects on the acoustic modes inside the vortex core,
and it has nothing to do with the energy cascade.
However in the present case, the magnetic field plays a role of driving force to 
create vortices in the BEC,
and therefore the interplay of the applied magnetic field and the random local noise
can produce nontrivial phenomena in the infrared regime.
As the system is 2D, it is interesting to see if the system exhibits some
specific phenomena like the inverse energy cascade as in the classical 
counterpart \cite{Kraichnan}
as well as some common feature of the general classical fluid mechanics \cite{Kolmogorov,Falkovich}.
On the other hand, by varying the strength of the applied magnetic field \cite{Zoller,Aidelsburger_Miyake},
the total vortex number, i.e., the enstrophy in the present case, can be controlled.
This kind of versatility is specific feature of the cold atom system, 
and we expect that new aspects of the vortex dynamics are observed by experiments
on the present cold-atom system.

For the white noise inducing local fluctuations of density, we consider a thermal 
noise caused by non-condensed ingredient of cold atoms or ideal thermal reservoir, 
which is expected to exist in any cold atomic gas systems.
Then our target model is {\em a 2D Bose-Hubbard model }(BHM){\em, in which
effects of the thermal noise and the concomitant dissipation via the 
fluctuation-dissipation theorem} (FDT){\em are taken into account}.
We study real-time dynamics and universal law of {\it a few} quantized vortices created 
in SF state by the noise and magnetic field.
The main purpose of the present work is to see the elementary process of 
vortex creation by the above external effects.
The thermal noise may affect intrinsically dynamical behavior of the BHM in a weak
magnetic field, which is realized in the recent experiments \cite{Zoller,Aidelsburger_Miyake}.
To study the phenomena, we employ a Langevin type of the Gross-Pitaevskii equation,
called a stochastic Gross-Pitaevskii equation (SGPE) \cite{Cockburn,Stoof,Gautam,Moon,Stoof2}.
This equation includes a noise term and a dissipative term, which are connected with
each other by the FDT \cite{Kamenev}.
As we show, these two terms considerably affect the vortex dynamics in the BEC.
In fact, we found that vortices are generated near the boundaries of the BEC and
they move into the bulk of the BEC.
In the bulk, the vortices exhibit a large cyclotron motion as a result of the inverse
cascade of energy.
The infrared regime of the incompressible energy spectrum (IES) of the kinetic energy 
exhibits the Kolmogorov $k^{-5/3}$ spectrum.
After some period, which we call {\em transition period},
a large cyclotron motion of vortices gradually gets weak and vortex configuration 
approaches to a steady state of simple circular motions.
Then, the IES tends to have $k^{-1}$ spectrum instead of $k^{-5/3}$. 
On the other hand for the ultra-violet regime of the wave number, we found that  
the IES always exhibits $k^{-3}$ power law
if vortices exist in the bulk of the system.

This paper is organized as follows. 
In Sec.II, we shall explain the BHM in the existence of a weak magnetic field,
the local random noise, and the dissipation.
Previous works \cite{Cockburn,Stoof,Stoof2,Kamenev} 
on the open quantum system obtained by integrating out
``external degrees of freedom" are quite useful to study the present target system. 
We show the SGPE of the BHM assuming 
that the noise and dissipation come from a thermal bath.
In Sec.III, we exhibit the numerical results obtained by solving the SGPE of the BHM.
We show that vortices intrude into the bulk SF through the boundary of the system.
Depending on the strength of the dissipation and noise, the number of intruding vortices
varies.
This phenomenon is a result of the inverse cascade of energy.
Intruding vortices exhibit a large cyclotron motion in the bulk SF.
We also measure the temporal behavior the IES of the kinetic energy and the velocity of the fastest intruding vortex. 
From these measurement,  we discuss the relationship between the power law of the IES and the velocity of the intruding vortex. 
Section IV is devoted for discussion and conclusion.
In this study, we consider the local noise stems from the thermal bath.
However in the ultra-cold systems, it is possible to generate an artificial 
local noise. We mention how to generate an artificial local noise in the
ultra-cold atomic systems in Sec.IV.

\section{Model and SGPE}

The Hamiltonian of the BHM in an artificial gauge field is given as \cite{Zoller}
\begin{eqnarray}
H_{\rm BH}&=&-J\sum_{i,j}(a^{\dagger}_{i}e^{-iA_{i,j}}a_{j}+{\rm h.c.})\nonumber\\
&&+\frac{U}{2}\sum_{i}n_{i}(n_{i}-1),
\label{MGBHM}
\end{eqnarray}
where $a_i \ (a^\dagger_i)$ is the boson annihilation (creation) operator at site $i$ and 
$n_{i}= a^{\dagger}_{i} a_{i}$. 
$\langle i,j \rangle$ denotes a pair of nearest-neighbor (NN) sites.
Also, $J$ is hopping strength, $U$ on-site interaction, and $n_{i}$ number operator of 
the boson.
These parameters are highly controllable in optical lattice system \cite{Bloch}.
The artificial gauge field $A_{i,j}$ is given in the {\it axial-gauge}: 
$A_{i,i+{\hat x}}=-2\pi f y$ and $A_{i,i+{\hat y}}= 0 \ (i=(x,y))$, which represents 
a uniform magnetic field 
and satisfies $\sum_{\rm plaquette} A_{i,j}=2\pi f$ with a parameter 
$f$ [$0\leq f \leq 1$], the strength of the magnetic field.
This gauge field can be experimentally generated by a Raman induced 
assisted tunneling in optical super-lattice \cite{Aidelsburger_Miyake}.
In this paper, we consider a BEC regime in a weak magnetic field. i.e.,
$J \gg  U$, and the flux parameter $f$ is small.  

We shall study the vortex dynamics in the BHM.
To this end, by replacing the field operator $a^{(\dagger)}_{i}$ with a classical 
condensate field $\psi_{i} \ (\psi^\ast)$, 
the lattice version of Gross-Pitaevskii equation (GPE) is given as follows \cite{Polkovnikov},
\begin{eqnarray}
i\hbar\frac{\partial \psi_{i}}{\partial t}=\biggl[ -\sum_{j}Je^{-iA_{i,j}}
\psi_{j}+U|\psi_{i}|^{2}\psi_{i}-\mu \psi_{i}\biggr],
\label{DGPE}
\end{eqnarray}
where a chemical potential term $\mu\sum_{i}n_{i}$ has been added to $H_{\rm BH}$.
In what follows, we shall consider the extended formulation of the GPE of the BHM
including the thermal noise and dissipation effects. 
This phenomenological model is given by a Langevin type formulation of the lattice
GPE.
For the GPE in the continuum space, see Refs.\cite{Kamenev,Stoof,Cockburn}.
Then, including the noise and dissipative terms, we have
\begin{eqnarray}
&&i\hbar\frac{\partial \psi_{i}}{\partial t}=-\sum_{j}J_{ij}\psi_{j}+U|\psi_{i}|^{2}\psi_{i}\nonumber\\
&&\;\;\;\;\; -\mu \psi_{i}+iR_{i}(t)\psi_{i}+\eta_{i}(t),
\label{SDGPE}\\
&&\langle \eta^{*}_{i}(t)\eta_{j}(t')\rangle = i\frac{\hbar ^{2}}{2}
\Sigma ^{K}_{i}(t)\delta (t-t')\delta_{i,j},
\label{WN}
\end{eqnarray}
where $J_{ij}\equiv Je^{-iA_{i,j}}$, $R_{i}(t)$ coming from dissipative coupling to 
a thermal bath, and $\eta_{i}(t)$ is the random field corresponding to the noise 
with the local correlation in Eq.(\ref{WN}).
$\Sigma ^{K}_{i}(t)$ is the lattice Keldysh self-energy, and
this self-energy is derived from the on-site interaction term 
by using the second-order perturbation theory \cite{Kamenev,Stoof}.
Here, we require that the above equation satisfies a fluctuation-dissipation theorem (FDT), 
which is given as \cite{Cockburn}
\begin{eqnarray}
&iR_{i}(t)\psi_{i}=-\frac{\hbar}{2}\Sigma ^{K}_{i}(t)\biggl[1+2\frac{1}{e^{\beta(\epsilon_{s}-\mu)}-1}\biggr]^{-1}\psi_{i}\nonumber\\
&\Rightarrow  iR_{i}(t)\psi_{i}\approx -\frac{\hbar}{4}\Sigma ^{K}_{i}(t)\beta(\epsilon_{s}-\mu)\psi_{i},\label{FDT}
\end{eqnarray}
where $\epsilon_{s}$ is a condensate energy for single particle derived from the  
Hamiltonian of the BHM, i.e., the Bogoliubov excitation spectrum, and 
$\beta=1/(k_{\rm B}T)$ with the Boltzmann constant $k_{\rm B}$.
To reformulate the GPE including the dissipation,
this single particle energy is changed to a time derivative operator as
$\epsilon_{s}\rightarrow i\hbar\partial_{t}$.
Then the FDT relation (\ref{FDT}) and the substitution of the energy spectrum $\epsilon_{s}$ give the following equation,
\begin{eqnarray}
&&\biggl[i\hbar-i\frac{\beta\hbar^{2}}{4}\Sigma^{K}_{i}(t)\biggr]\frac{\partial \psi_{i}}{\partial t}\nonumber\\
&&=-\sum_{j}J_{ij}\psi_{j}+U|\psi_{i}|^{2}\psi_{i}-(\mu-\mu_{{\rm eff}}) \psi_{i}+\eta_{i}(t),\label{SGPE1}\\
&&\mu_{{\rm eff}}=\frac{\beta\hbar}{4}\Sigma^{K}_{i}(t)\mu.
\end{eqnarray}
In the later practical numerical calculation,
we treat the Keldysh self-energy as an effective parameter independent 
of the space-time coordinate,
$i\frac{\beta\hbar}{4}\Sigma^{K}_{i}(t)\rightarrow \gamma_d$.
With this approximation, the SGPE is given as
\begin{eqnarray}
\hbar(i-\gamma_{d})\frac{\partial \psi_{i}}{\partial t}
=-\sum_{j}J_{i,j}\psi_{j}+U|\psi_{i}|^{2}\psi_{i}-\mu ' \psi_{i}+\eta_{i}(t),\nonumber\\
\label{SGPE3}
\end{eqnarray}
where $\mu'\equiv \mu-\mu_{\rm eff}$.
And then the noise correlation function is 
given by 
\begin{eqnarray}
\langle \eta^{*}_{i}(t)\eta_{j}(t')\rangle = 2\gamma_{d}\hbar k_{B}T\delta (t-t')\delta_{i,j}.
\label{noise1}
\end{eqnarray}
This noise is nothing but the genuine white noise

\begin{figure}[t]
\centering
\begin{center}
\includegraphics[width=8cm]{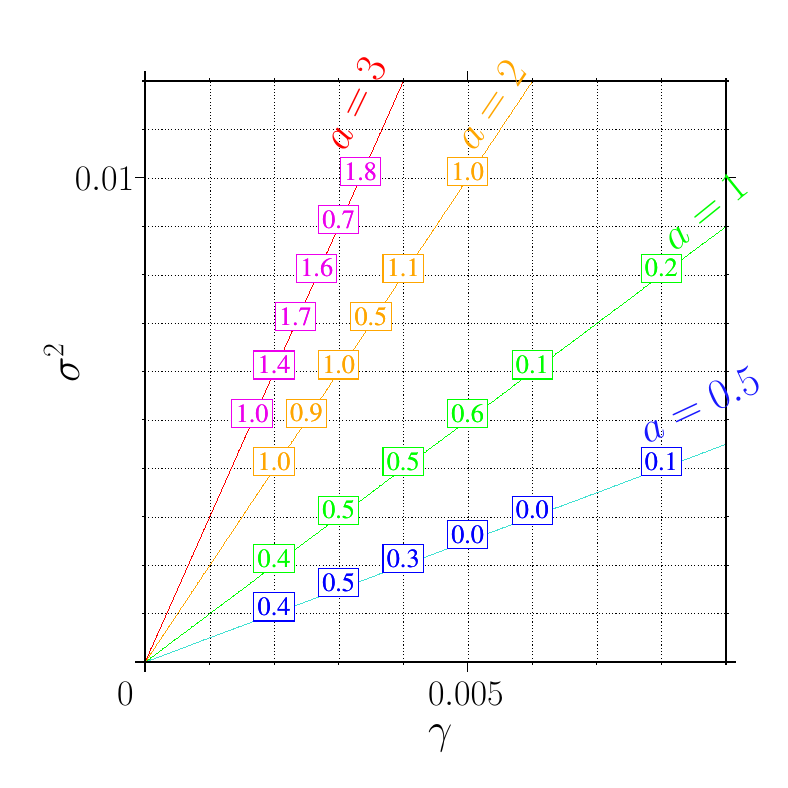}
\end{center}
\caption{(Color online) Average vortex number in the $(\sigma^{2}$-$\gamma_{d})$ plane.
We take 10 samples in each point by varying the white noise.
In finite number of votex cases, 
shortly after switching on the noise field $\xi_{i}$ vortex 
starts to intrude the bulk superfluid with rotational motion.
Each point indicates steady states after a transient time (almost $t' \sim O(10^5)$).
$a\equiv\sigma^{2}/\gamma_{d}$}
\label{gamma_sigma}
\end{figure}
\begin{figure*}[t]
\includegraphics[width=12cm]{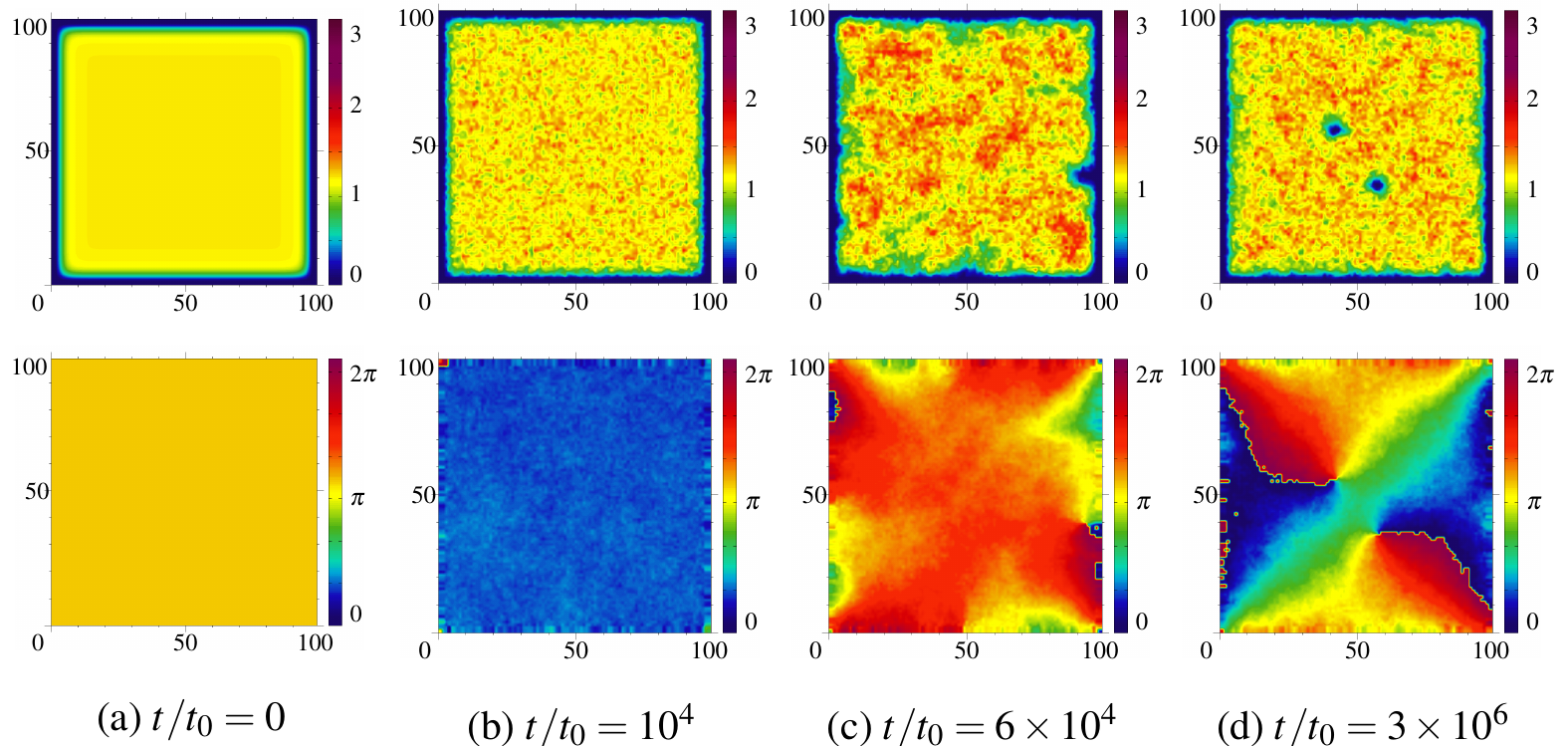}
\caption{(Color online) Snapshots of atomic density and phase of the condensed
field for two-vortex configuration at $a=3$ in Fig. \ref{gamma_sigma}.
We first turn on the noise and then the magnetic field.
The noise makes the density fluctuations and the magnetic field
create vortices at the boundary.
Vortices move into the bulk BEC from the boundary.
 $\sigma^{2}=0.006$ and $\gamma_{d}=0.002$. }
\label{density_phase}
\end{figure*}

Before performing the numerical simulations, we rescale the parameters and fields in 
the SGPE in Eq.(\ref{SGPE3}).
We first introduce the density unit $\rho_{0}$ and time unit $t_{0}$, and then the time $t$ 
and the field $\psi_{i}$ are scaled as $t=t_{0}{\tilde t}$ and $\psi_{i}=\sqrt{\rho_{0}}\phi_{i}$. 
We divide Eq.(\ref{SGPE3}) by $\sqrt{\rho_{0}}$,
and take the hopping parameter $J$ as the units of the energy. 
Then the SGPE is rewritten as,
\begin{eqnarray}
\hbar(i-\gamma_{d})\frac{1}{Jt_{0}}\frac{\partial \phi_{i}}{\partial {\tilde t}}
&=&-\sum_{j}e^{-iA_{i,j}}\phi_{j}+U'|\phi_{i}|^{2}\phi_{i}\nonumber\\
&&-\mu '' \phi_{i}+\frac{1}{J\sqrt{\rho_{0}}}\eta_{i}({\tilde t}),
\label{SGPE5}
\end{eqnarray} 
where $U'=\rho_{0}U/J$, and $\mu''=\mu'/J$. 
We introduce a rescaled noise field 
$\xi_{i}({\tilde t})\equiv (1/(J\sqrt{\rho_{0}}))\eta_{i}({\tilde t})$, whose
correlation is given by,
\begin{eqnarray}
\langle \xi^{*}_{i}({\tilde t})\xi_{j}({\tilde t}')\rangle &=& \frac{2\gamma_{d}\hbar k_{B}T}{J\rho_{0}}\delta ({\tilde t}-{\tilde t}')\delta_{i,j} \nonumber \\
&=&2\sigma^2\delta ({\tilde t}-{\tilde t}')\delta_{i,j},
\label{noise2}
\end{eqnarray} 
Here we have put $t_0\equiv\hbar/J$, and $\sigma^2\equiv \frac{\gamma_{d}\hbar k_{B}T}{J^{2}\rho_{0}}$.
From this treatment, the following dimensionless SGPE are obtained,  
\begin{eqnarray}
(i -\gamma_{d})\frac{\partial \phi_{i}}{\partial {\tilde t}}
&=&-\sum_{j}e^{-iA_{i,j}}\phi_{j}\nonumber\\
&+&U'|\phi_{i}|^{2}\phi_{i}-\mu '' \phi_{i}+\xi_{i}({\tilde t}).
\label{SGPE_nd}
\end{eqnarray}

Let us estimate typical values of the units.
In BEC regime, the typical hopping energy 
$J/\hbar\sim O (10^2 \sim 10^3$ [${\rm Hz}$]) \cite{Bloch,Endres} ,  
then from the scaling relation $\hbar/(t_{0}J)=1$, the time unit becomes 
$t_0\sim 2.0$ [msec].
Also we put the density $\rho_{0}=1$, i.e., one atom per one lattice site,
which is feasible in real experiments.
However, it is obvious that the later numerical results are applicable
for other values of $\rho_0$ by rescaling $\sigma$ and $U'$, i.e.,
$(\sigma, U')\rightarrow (\sqrt{\rho_0}\sigma, U'/\rho_0)$.

Finally, we remark that there exist constraints 
on temperature at which the phenomena predicted in the present work
are to be observed in experiments.
The first constraint comes from the fact that only the lowest-band in the 
optical lattice is considered in the present study.
Since we ignore the effects of the excited bands,
the temperature is limited by $k_{B}T\leq \omega_{0}\hbar$, where 
$\omega_{0}\hbar$ is a band gap to the first-excited band. 
This band gap can be calculated by the band calculation for
the optical lattice potential. 
As an example, for  ${}^{87}$Rb gases in a typical square 
optical lattice \cite{Bloch}, 
the band gap $\omega_{0}\hbar$ is estimated as $4 - 16$KHz ($200 - 800$nK) 
for typical lattice depth $(4 - 21){\rm E_{r}}$, where 
${\rm E_{r}}$ is the recoil enegy, and for the ${}^{87}{\rm Rb}$,
${\rm E_{r}}=2$KHz ($\sim 100$ nK) \cite{Endres}.
Second, the system temperature has to be lower than the SF transition
temperature $T_c$.
While it is expected to be $T_{c} \sim 6J/{\rm k_{B}}$ \cite{Bloch,Trotzky} 
in large $U/J$ regime,  
the critical temperature $T_{c}$ in  a deep BEC regime (i.e., $J/U$ is large) at unit filling  
is not precisely known. 
Therefore, we simply assume that the system temperature $T$ is  $T\leq 200$nK.

\begin{figure*}[t]
\includegraphics[width=12cm]{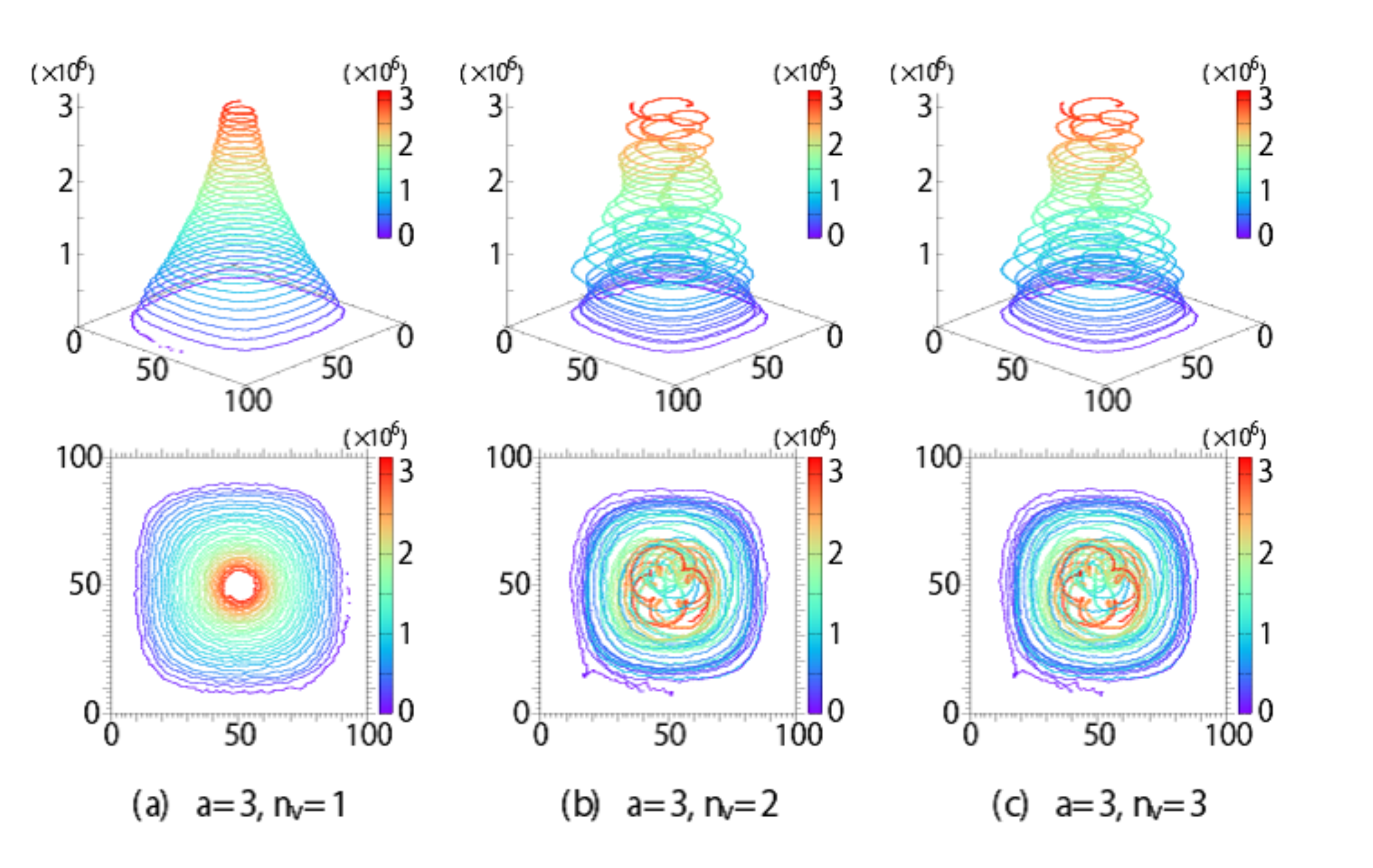}
\caption{(Color online) 
Real-time dynamics of vortices obtained by solving the SGPE. 
In the case that energy injection from the noise field $\xi_{i}$ and dissipation are 
balanced by the FDT, 
density fluctuation caused by the noise $\xi_{i}$ creates quantized vortex in the system boundary and then vortex circular motion appears. 
Multiple intruding vortices tend to exhibit cluster-like rotational motion.
The length scale of the radius of rotational vortex motion  is larger than the scale of the energy injection through the noise field.
$n_v$ is the number of vortices.}
\label{trajectory1}
\end{figure*}

\section{Numerical results: BHM coupled with magnetic field and white noise}

In this section, we show the results of the numerical study on the BHM, in which 
effects of the white noise and magnetic field are taken into account. 
In the numerical simulations, we fix the values of
the parameters $\sigma^{2}$ and $\gamma_{d}$, and
study phenomena to obtain a ``phase diagram"in the $(\sigma^{2}-\gamma_{d})$-plane. 
We solve the SGPE of Eq.~(\ref{SGPE_nd}) by the Crank-Nicolson implicit scheme 
\cite{Crank_Nicolson} for the 2D square lattice system with $(100\times 100)$ sites.
We employ the sharp-boundary condition by introducing a wall potential with 
the height $100J$, which corresponds to the box-trap 
system realized in the recent experiments \cite{box_trap}.   
We set the parameters as $U'=0.2$, $f=0.001381$ with $\rho_{0}=1$, which
corresponds to a deep SF region in a weak magnetic field.
As initial states, we prepare a homogeneous SF state 
in which the phase and density of $\{\phi_i\}$ are almost uniform.
In our numerics, starting with this initial state,the random noise $\{\xi_{i}\}$ is turned on at $t'=0$, 
and then afterward at $t'=10^3$,  the uniform magnetic field $A_{i,j}$ is turned on.
We employ the canonical ensemble by controlling the chemical potential
$\mu''$ suitably along the time evolution, 
i.e., the total particle number is conserved along the time evolution although there
is the dissipation.

Before going into study on the target system, we examined the BHM coupled 
with a weak magnetic field in the absence of the noise.
We found that {\em vortices are not generated in this case}.
This is nothing but the Meissner effect.
Vortices are generated after application of the noise to this system.
This means that the interplay of the magnetic field and the local noise
is an essential ingredient for generation of bulk vortices in the present system.
In addition, we examined the result of the reverse process,
 i.e., random noise is turned on first and then
the magnetic field is applied.
We observed that the random noise makes phases of $\{\phi_i\}$ uniform,
i.e., the SF order is enhanced by the random noise.
This is a kind of {\em order by disorder}, and is interesting phenomenon.
We shall study this phenomenon in detail in the near future.

\subsection{Number of intruding vortices and their behavior}

In this and subsequent sections, we shall show the numerical results for
the model introduced in Sec.II.
First in Fig.~\ref{gamma_sigma}, we plot the average number of vortices generated 
by the noise effect in the $(\sigma^{2}$-$\gamma_{d})$-plane. 
We examined ten samples of the random noise for each point in Fig.~\ref{gamma_sigma}.
At each point, the configuration approaches to an almost steady state 
after a transition period (about $t' \sim O(10^5)$).  
Typical evolution of the system is as follows.
When the noise and magnetic field are turned on, vortices start to intrude into 
the bulk SF from the boundary of the system, and then the intruding vortex exhibits 
a sightly complicated cyclotron motion and reaches to a final steady state of a 
circular motion. 
This phenomenon is caused by the interplay between the external magnetic field and
the white noise as we explained above.
The resultant dynamical behavior is highly sensitive to the parameters $\sigma^{2}$ 
and $\gamma_{d}$, 
i.e., the number of intruding vortex varies depending on the parameters $\sigma^{2}$ 
and $\gamma_{d}$.
As shown in Fig.~\ref{gamma_sigma}, the number of intruding vortex tends to large 
as the ratio $a$ defined by $a=\sigma^{2}/\gamma_{d}$ is getting large.
The reason is that the large noise amplitude ($\sigma^2$: large) directly modulates 
the superfluid density, and then topological defects in the SF tend to be created in the boundary of the system, thus vortices tend to form. 
Time evolution of the density and phase of the BEC
for the two-vortex case, $\sigma^{2}=0.006$ and $\gamma_{d}=0.002$, in 
Fig.~\ref{gamma_sigma} is shown in Fig.~\ref{density_phase}. 
The results in the upper panels show that the density of the BEC is modulated 
by the noise and holes, which generate in the boundary, move into the bulk of the BEC.
The lower panels show that the BEC has a phase coherence but the singularities
exist at the locations of hole.
These behaviors of the density and phase clearly indicate the generation of
two vortices and their behavior in the BEC.

Figure~\ref{trajectory1} shows the detailed trajectories of the intruding vortices 
obtained by solving the SGPE.
In the practical numerical study, we first introduce the noise 
and then apply the magnetic field.
Each trajectory in Fig.~\ref{trajectory1} corresponds to a point indicated
in Fig.\ref{gamma_sigma}.
In all observed cases in Fig.~\ref{gamma_sigma} except the cases without vortices, 
the BEC density starts to fluctuate immediately after turning on the noise,
and vortices start to intrude from the boundary of system into the bulk. 
Then the spiral motion of vortices takes place until they reach to a stable circular
motion.
We note that these vortex motions are related to 
{\it inverse energy cascade} phenomena. 
Some previous works \cite{Bradley1,Bradley2,Bradley3,Bradley4,Skaugen} reported 
that such a phenomenon is a typical behavior of quantum turbulence in 
2D SF in contrast to quantum turbulence in 3D systems \cite{Kobayashi1}.
In the present SGPE, energy injection is performed by the noise field $\{\xi_j\}$. 
As Eq.~(\ref{noise2}) shows, the energy injection in the present system has the
very small spatial-scale and short time-scale properties. 
On the other hand, the rotational vortex motion  in Fig.~\ref{trajectory1} is 
a large scale behavior and therefore the observed 
phenomenon is a kind of the inverse energy cascade. 
In later section, we shall calculate the IES of the system including vortices.
The result reveals an interesting relationship between the vortex motion and
the IES.

\subsection{Incompressible kinetic energy spectra}
\begin{figure}[t]
\includegraphics[width=8cm]{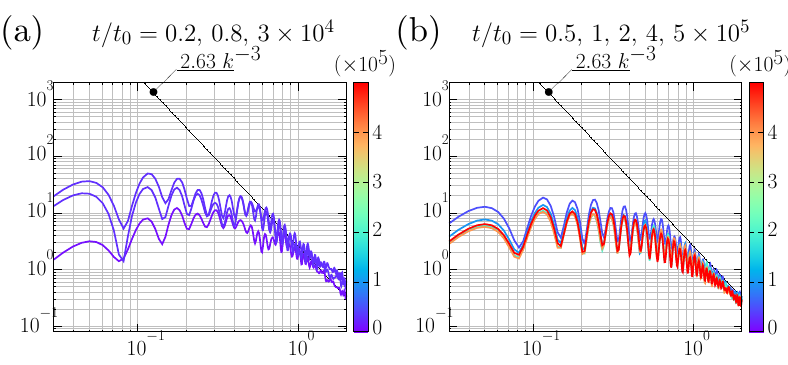}
\caption{(Color online)  Incompressible kinetic energy spectra (IES) for no intruding vortex case ($\sigma^{2}=0$).
(a) The IESs in $0<t/t_{0}< 2\times 10^5$. The intruding vortex appears from the boundary of the system. 
The IES exhibits the $k^{-3}$ power law in the ultraviolet regime $k>2\pi\zeta^{-1}$. 
 (b) The IESs in $2\times 10^5 <t/t_{0}$. The system has no vortex in the bulk SF. 
The IES has no power law.}
\label{IES0}
\end{figure}
\begin{figure*}[t]
\includegraphics[width=14cm]{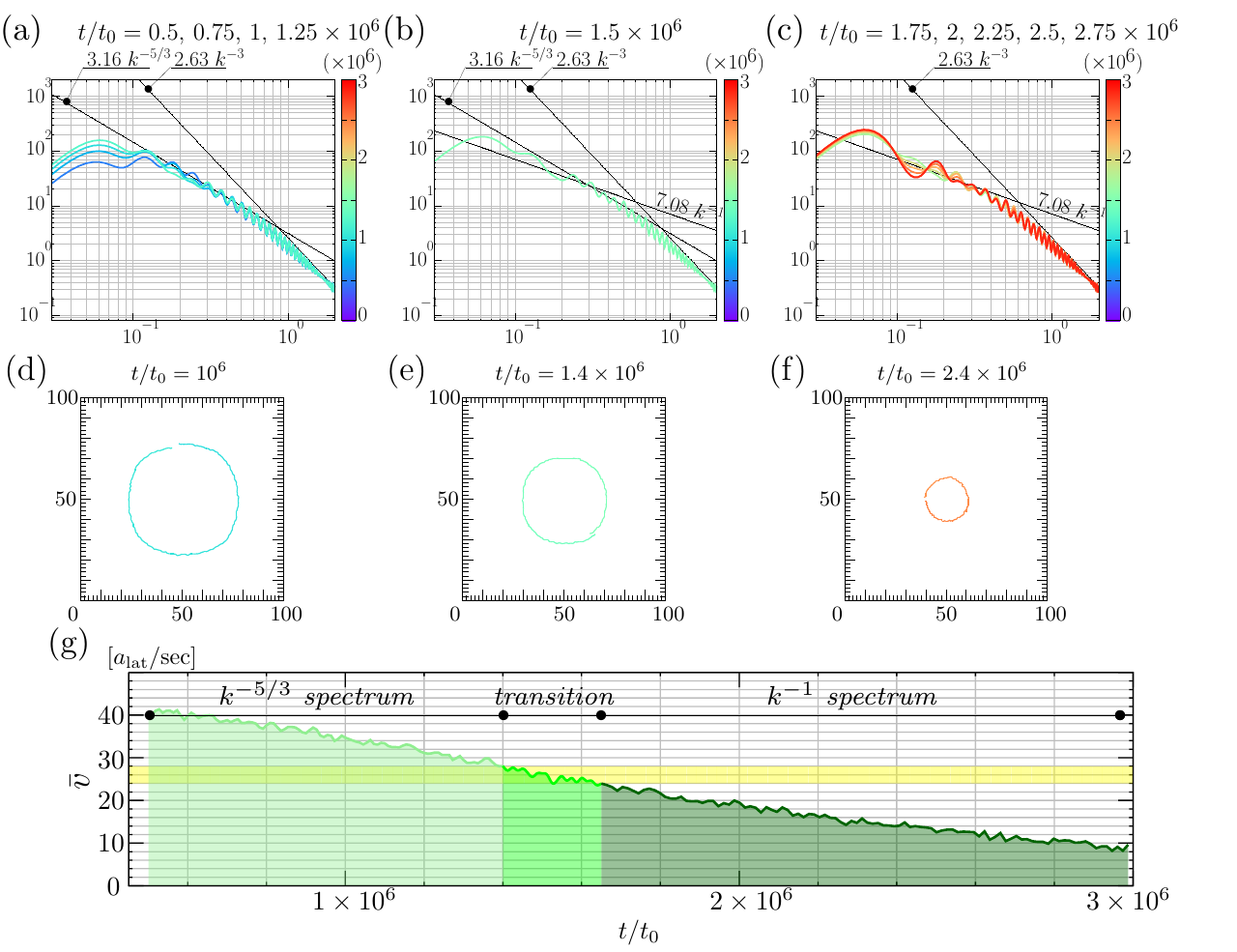}
\caption{(Color online) Incompressible kinetic energy spectra and the vortex velocity 
${\bar v}(t)$ for one intruding vortex case ($a=3$). 
(a)-(c): The IESs for each time. (d)-(f): The typical one-cycle trajectories of vortex. 
(g): The velocity of the vortex ${\bar v}(t)$. 
In the cases of the large cyclotron motion of vortices for ${\bar v}(t)> 30$, the IES (a) exhibits the inverse energy cascade 
and the Kolgomorov $k^{-5/3}$ spectrum in the inertial
range, the momentum space $k<2\pi\zeta^{-1}$.
On the other hand, in the IESs in (b) and (c)  where the large cyclotron motion of vortices are suppressed, ${\bar v}(t)< 30$, 
the Kolmogorov $k^{-5/3}$ spectrum in the infrared regime $(L\zeta)^{-1}<k<2\pi\zeta^{-1}$ is replaced with $k^{-1}$ spectrum.}
\label{IES1}
\end{figure*}

Properties of dynamical flow of vortices in the present compressible system is
characterized by measuring a kinetic energy spectrum and identifying its power law. 
In particular, we focus on the energy spectrum of the incompressible part of 
the kinetic energy of the vortex system. 
The SF velocity field $u_{i,\mu}$ ($\mu=1(2)$ represents $x$ ($y$)-direction) 
is given as follows by using the parameterization $\phi_{i}=\sqrt{\rho_{i}}e^{i\theta_{i}}$
($\rho_{i}$ is the superfluid density and $\theta_{i}$ is the phase), 
$u_{i,\mu}=\sqrt{\rho_{i}}\nabla_\mu \theta_{i}$ 
(the operator $\nabla_{\mu}$ is the difference operator, 
$\nabla_{\mu}f_{i}\equiv f_{i+\mu}-f_{i}$). 
As the practical calculation shows, the SF density $\rho_{i}$ is smooth over 
the whole system, and therefore we put $\nabla_{\mu}\rho_{i}\approx  0$. 
Next, the field $u_{i,\mu}$ is decomposed into compressible part $u^{C}_{i,\mu}$ 
and incompressible part $u^{I}_{i,\mu}$ using the Helmholtz theorem,
$u_{i,\mu}=u^{C}_{i,\mu}+u^{I}_{i,\mu}$, 
where $\sum_{\mu, \nu}\epsilon_{\mu\nu}\nabla_{\mu}u^{C}_{i,\nu}=0 \ 
[\epsilon_{12}=-\epsilon_{21}=1, \epsilon_{11}=\epsilon_{22}=0]$ 
and $\sum_{\mu}\nabla_{\mu}u^{I}_{i,\mu}=0$. 
In the practical calculation,
we obtain the incompressible part $u^{I}_{i,\mu}$ by using the vector 
potential ${\bf W}_{i}=(0,0,W^{z}_{i})$, which is 
obtained by solving the coalition difference equations for the divergence free condition, $\sum_{\mu}\nabla_{\mu}u^{I}_{i,\mu}=0$. 
The IES is then given as,
\begin{eqnarray}
E_{I}(k)&=&\int_{|{\bf k}|=k\in {\rm 1st B.Z.}}dk\biggl[\tilde{u}^{I}_{\mu}({\bf k})\cdot \tilde{u}^{I}_{\mu}(-{\bf k})  \biggr],\label{EIn}\\
\tilde{u}^{I}_{\mu}({\bf k})&=&\frac{1}{2\pi}\sum_{i,\nu}\sum_{{\bf R}_i}e^{i{\bf k}
\cdot {\bf R}_{i}}\epsilon_{\mu\nu}\nabla _{\nu}W^{z}_{i}\label{Fu}.
\end{eqnarray}  
Here, the momentum vector ${\bf k}$ is restricted to the first brillouin zone (B.Z.). 
The $E_{I}(k)$ measures the energy carried by vortices. 

In order to study the relation 
between the IES and the instantaneous velocity of vortex, we define
the vortex velocity at time $t$, ${\bar v}(t)$, as follows,
\begin{eqnarray}
{\bar v}(t)&=&\frac{1}{10\Delta T}\sum^{9}_{n=0}d_{n}(t),\label{velocity}\\
d_{n}(t)&=&|{\bf r}(t+(n+1)\Delta T)-{\bf r}(t+n\Delta T)|, \label{distance}
\end{eqnarray}
where ${\bf r}(t)$ is the coordinate of the core of the vortex,
and therefore
$d_{n}(t)$ obviously represents  the difference of the distance in 
the small time interval $\Delta T$.
To minimize numerical errors, the velocity ${\bar v}(t)$ is calculated 
by taking an average over ten samples of the adjacent times around $t$.
In our numerics, we take $\Delta T=2.5\times 10^3$ and 
focus on the velocity ${\bar v}(t)$ of the fastest vortex intruding into the bulk SF.

\begin{figure*}[t]
\includegraphics[width=14cm]{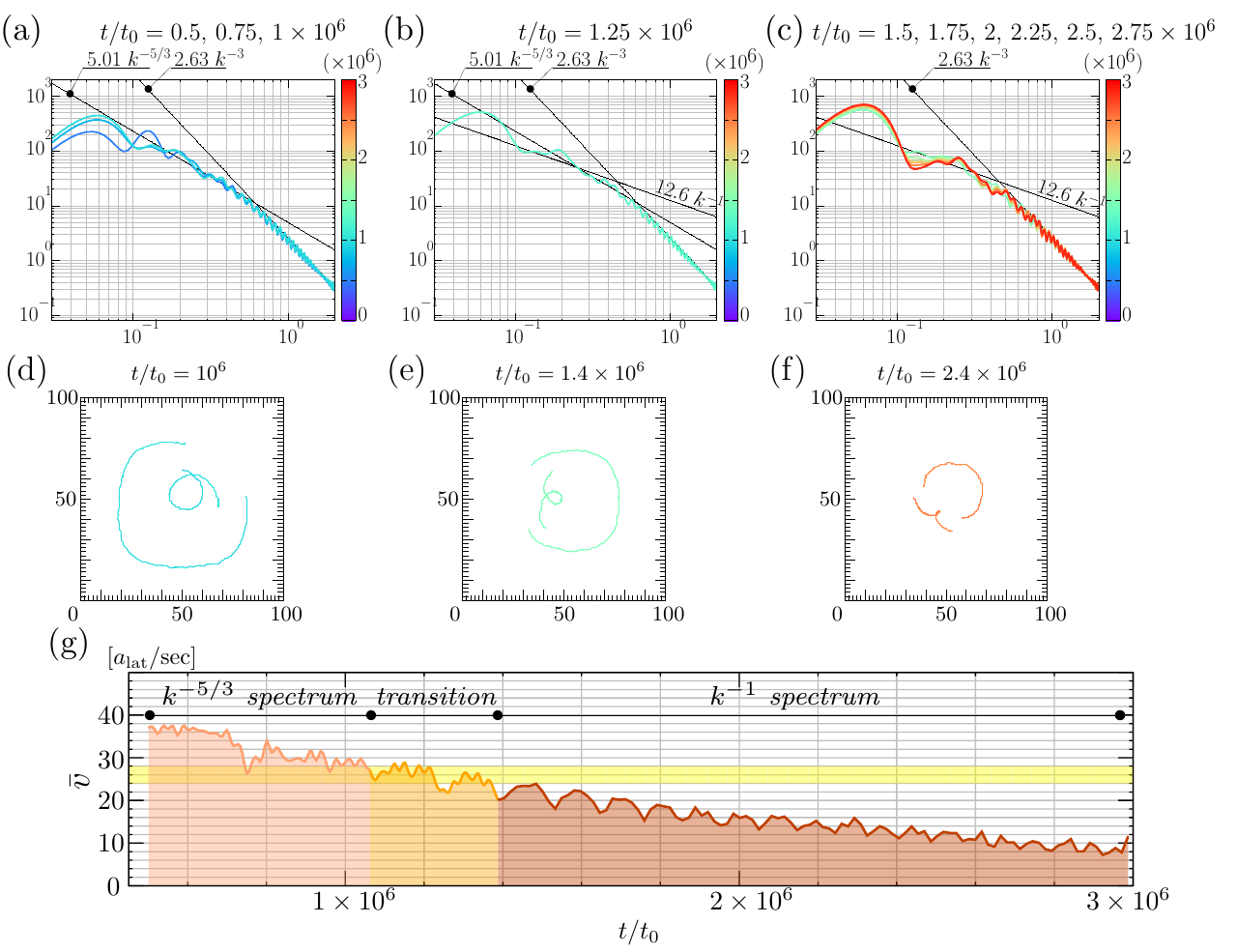}
\caption{(Color online) Incompressible kinetic energy spectra and 
the vortex velocity ${\bar v}(t)$ for two intruding vortex case ($a=3$). 
(a)-(c): The IESs for each time. (d)-(f): 
The typical one-cycle trajectories of vortex around a time $t$. 
(g): The velocity of the fastest vortex ${\bar v}(t)$. 
We observe the same behavior of the IESs and ${\bar v}(t)$ with the 
one-vortex case.}
\label{IES2}
\end{figure*}
\begin{figure*}[t]
\includegraphics[width=14cm]{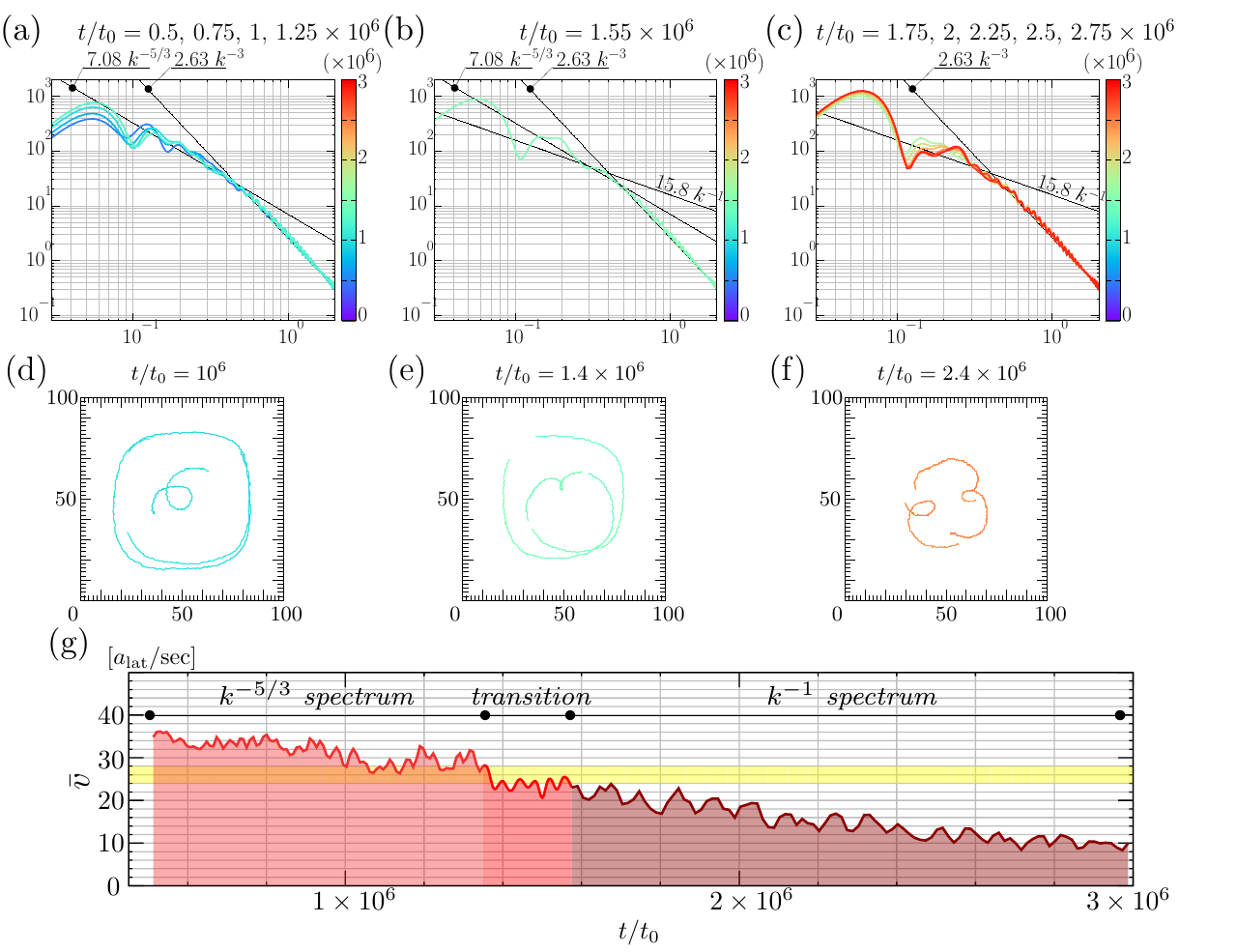}
\caption{(Color online) Incompressible kinetic energy spectra and the velocity 
${\bar v}(t)$ for three intruding vortex case ($a=3$). 
(a)-(c): The IESs for each time. (d)-(f): The typical one-cycle trajectories of vortices. 
(g): The velocity of the fastest vortex ${\bar v}(t)$.
We observe the same behavior of the IESs and ${\bar v}(t)$ with the 
one-vortex case}
\label{IES3}
\end{figure*}

First, the calculated IESs for the vortexless case are shown in 
Fig.~\ref{IES0} (a) and (b).
There the noise is absent whereas the weak magnetic field exists.
Once the magnetic field is applied, a few intruding vortices
usually appear on the boundary of the system.
The IESs of such a situation are shown in Fig.~\ref{IES0} (a). 
Here, we found that 
the IES of the bulk SF has the $k^{-3}$ power-law behavior in the ultraviolet regime $k>2\pi\zeta^{-1}$, 
where $\zeta$ is the healing length, i.e., the radius of vortex core. 
The healing length $\zeta$ is estimated 
as $\zeta \sim 5a$ for the weak magnetic field $f=0.001381$.
It is expected that this power law comes from the acoustic mode of 
the vortex core  \cite{Bradley1,Bradley2}. 
Our simulation shows that, after a short period, all the intruding vortices move out 
the bulk SF, and then the system is stabilized to the final steady state without 
intruding vortices. 
The IESs of the steady are shown in Fig.~\ref{IES0} (b). 
The result shows that the IES does not exhibit the $k^{-3}$ power law in 
the whole $k$-region in the final steady state.
This behavior of the IES in Fig.~\ref{IES0} (a) and (b) is understood as follows.
Just after turning on the magnetic field, 
vortices first intrude into the system, and as a result, 
the $k^{-3}$ power law of the IES appears.
However, the vortices leave the bulk BEC within a short period,
and then the $k^{-3}$ power-law disappears.
As the local noise is absent in the present case, the SF tends to approach to 
the lowest-energy state without vortices as a result of the dissipation.

Next, numerical results of the IES and 
the velocity of the vortex ${\bar v}(t)$ are shown for 
each real-time dynamics in Fig.~\ref{trajectory1} (a)-(c).
See Figs.~\ref{IES1}-\ref{IES3}, which correspond to Fig.~\ref{trajectory1} (a)-(c),
respectively. 
In the upper panels of Figs.~\ref{IES1}-\ref{IES3}, we show the time evolution
of the IES.
In the middle panels, typical motions of votices are exhibited, and in the lower panels
the velocity of the vortex ${\bar v}(t)$ is shown as a function of time. 
From these result, we found general relationship between the IES and the velocity of 
the fastest vortex:  
When the velocity ${\bar v}(t)$ is larger than a critical point ${\bar v}(t)\sim 28$ 
[$a_{lat}$/sec] ($a_{lat}$ is an optical lattice spacing, $\sim 500$[nm]), 
the IES exhibits the Kolmogorov $k^{-5/3}$ spectrum 
in the infrared regime $(L\zeta)^{-1}<k<2\pi\zeta^{-1}$ \cite{Kolmogorov,Kraichnan}
(the parameter $L$ is simply determined by the system size). 
This infrared regime is called inertial region.
We shall expect that the Kolmogorov $k^{-5/3}$ spectrum comes from 
the cyclotron motion of vortex. 
Let us see the detailed results in Figs.~\ref{IES1}-\ref{IES3}.

Figure \ref{IES1} shows the IES, typical trajectories of vortex and the velocity of the vortex, 
corresponding to the one-vortex configuration in Fig.~\ref{trajectory1} (a).
As seen from Fig.\ref{IES1} (a) and (g), in the time interval $0<t/t_{0}< 1.25\times 10^6$, 
the vortex velocity $\bar{v}(t)>28$ and 
the IES has the Kolmogorov $k^{-5/3}$ spectrum. 
The corresponding one-cycle trajectory of the vortex is shown in Fig.~\ref{IES1} (d), 
which exhibits that the vortex cyclotron-moves with a large velocity.
The above behaviors are generated by the interplay of the magnetic field and the noise. 
On the other hand below ${\bar v}(t)\sim 28$ at $t/t_0\sim 1.25\times 10^6$, 
the IES in Fig \ref{IES1} (b) deviates from the Kolmogorov $k^{-5/3}$ law and 
instead it fit a $k^{-1}$ power law in the infrared regime $(L\zeta)^{-1}<k<2\pi\zeta^{-1}$. 
As the velocity $\bar{v} (t)$ is decreasing furthermore, 
as seen from Fig \ref{IES1} (c), the IES spectrum 
seems to fit a $k^{-1}$ spectrum including somewhat oscillating behavior 
in the infrared regime $(L\zeta)^{-1}<k<2\pi\zeta^{-1}$ \cite{systemsize}.
The corresponding one-cycle trajectory of the vortex is shown in Fig.~\ref{IES1} (f).
The radius of the cyclotron motion is obviously smaller than that in the case of 
${\bar v}(t)> 28$.   

Systems with two or three intruding vortices have almost the same properties 
with the one with single vortex. 
The obtained IESs, typical trajectories of vortex and velocities of vortex are 
shown in Figs.~\ref{IES2} and \ref{IES3} that correspond to the configurations
in Fig.~\ref{trajectory1} (b) and (c).
The obtained results are quantitatively similar to the ones for single intruding vortex 
case shown in Fig.~\ref{IES1}.
In large velocity regime ${\bar v}(t) > 30$ in $t/t_{0}<1.4\times 10^{6}$, the results 
in Figs.~\ref{IES2} and \ref{IES3} show that the vortices exhibit somewhat large 
cyclotron motion.
Furthermore, the one-cycle trajectories of the vortices in Figs.~\ref{IES2}(d) and 
\ref{IES3}(d) exhibit clustering-like movements.  
The corresponding IES has the Kolmogorov $k^{-5/3}$ spectrum in the infrared regime 
$(L\zeta)^{-1}<k<2\pi\zeta^{-1}$, shown in Figs.~\ref{IES2} (a) and \ref{IES3} (a). 
Similar result was reported in Refs.\cite{Bradley2,Bradley3,Bradley4,Skaugen}, 
where the dynamics of many clustering vortices creates the Kolmogorov spectrum
as a result of the energy injection at scales $k \sim\zeta^{-1}$. 
Comparing with the above previous works, our system has only a few vortices, 
but the IES has the same Kolmogorov $k^{-5/3}$ spectrum in the infrared regime. 
Furthermore, it is believed that the energy injection at length scales smaller than 
the healing length $\zeta$ (i.e., $k>\zeta^{-1}$) cannot generate the inverse
energy-cascade phenomenon.
Therefore at first glance, the preset result seems to contradict the previous observation.
However in the present study, the weak but finite magnetic field is applied to
the system, and the interplay of the magnetic field and the local random noise
create the turbulence like behavior of the IES, i.e., the Kolmogorov spectrum. 
Therefore, there exist no contradictions between the previous and the present
observations, and the results obtained in the present work are a new finding.

As shown in Figs.~\ref{IES2}(c) and \ref{IES3}(c),
the IES exhibits the $k^{-1}$ spectrum instead of the $k^{-5/3}$
as the value of  the velocity $\bar{v}(t)$ decreases below the critical value
$\bar{v}(t)\sim 28$.
It should be remarked that the critical velocity is universal, i.e., it is independent of
the number of vortex in the bulk of the SF.
It should be also noticed that the oscillating behavior of the IES in the 
infrared region is getting large as the number of the bulk vortex increases.

Before closing this section, we summarize the nucleation process of vortex 
and inverse energy cascade phenomenon observed in this section.
Numerical study of the IES for the real-time dynamics of a finite number of vortex 
in Fig.~\ref{trajectory1} clearly shows that 
the IES is closely related to the behavior of the intruding vortices.
At first, the injected energy by the random noise induces the vortices at the
boundary of the system, in particular, the energy injected by the noise 
is used to creation of vortex cores.
Therefore, there does not exist the inertial region with $k^{-5/3}$ in the initial stage.
The ultraviolet region with $k^{-3}$ in the IES appears as the vortex cores develop. 
Then after the creation of vortex cores, the injected energy by the local 
noise co-operates with the external magnetic field
and creates the large movement of the vortices and also the inertial regime with 
the Kolmogorov $k^{-5/3}$ spectrum in the IES.
After the transition period, the dissipation of the system makes vortices stay 
a stable and steady state, and the IES exhibits the $k^{-1}$ spectrum as 
in the static single vortex configuration \cite{Bradley1}.


\section{Conclusion}

In this paper, we studied the 2D BHM that describes cold atomic gas 
coupled with an artificial magnetic field.
The model also takes into account effects of the dissipation and local white noise 
that are connected with each other by the FDT.
We applied the SGPE to the above BHM, and studied
the dynamical properties, in particular,  behavior of vortices by 
numerically solving the SGPE of the Langevin type equation. 

Here we mention another feasible physical system described in terms of the
above 2D BHM.
In this work, we assume that the dissipation and noise stem from  
the thermal fluctuations caused by the non-condensed atoms.
However, there is a possibility to create artificial noises for atomic density
by using optical speckle potentials in optical lattices \cite{Sanchez}.
This method may create various noises with some spacial and/or time length correlation,
and as a result, there appear a dissipation through the FDT.

The numerical simulations in this work showed the vortex-nucleation process 
induced by the interplay of the magnetic field and the local white noise in certain
parameter regime of the dissipation $\gamma_{d}$ and noise amplitude $\sigma^2$.
Moreover the SGPE simulation observed the intruding behavior of vortices. 
Vortices exhibit large cyclotron motions at early stage of the intrusion and
this behavior is understood as a phenomenon of the inverse energy cascade,
which is one of the characteristic elements of 2D turbulence.
To investigate the vortex dynamics in detail, we studied the time evolution of 
the IES and the velocity of vortcies in the bulk SF.
Once the intruding vortices appear in the bulk SF, the IES has the $k^{-3}$ power law 
in the ultraviolet regime $k>2\pi\zeta^{-1}$. 
Furthermore, the numerical calculation of the IES indicates that when the intruding vortices has large velocity and large cyclotron motion, 
the Kolgomolov $k^{-5/3}$ spectrum
appears in the infrared regime $k<2\pi\zeta^{-1}$, which corresponds to the inertial 
range of turbulence.
After the transition period, the large velocity and the large cyclotron motion are suppressed by the dissipation, 
and the Kolgomolov $k^{-5/3}$ spectrum in the IES
is replaced with the $k^{-1}$ spectrum. 
We found that  the above mentioned phenomena are independent to the number of 
the vortex in the SF.
We also found that there exists a critical velocity of the fastest vortex at
which the IES in the infrared regime changes from the Kolgomolov $k^{-5/3}$ spectrum 
to the $k^{-1}$ spectrum.  In appendix, as a subsidiary work, we report the study of cold atomic gases in 
a 2D optical lattice by means of the {\em continuum SGPE}.
Similar qualitative properties of vortices in BEC have been obtained there.

The detailed study on the origin of the Kolgomolov spectrum 
and the energy flux will be reported in our future work. 
Finally, we hope that optical lattice experiments observe such vortex dynamics 
numerical-simulated in this work.


\acknowledgments
We acknowledge K. Kikuchi and N. Shirasaki for useful discussion on the numerical
study of the SGPE. 
Y. K. acknowledges the support of a Grant-in-Aid for JSPS
Fellows (No.JP15J07370). This work was partially supported by Grant-in-Aid
for Scientific Research from Japan Society for the 
Promotion of Science under Grant No.JP26400246.

\appendix
\renewcommand{\thefigure}{\Alph{section}.\arabic{figure}}
\setcounter{figure}{0}
\renewcommand{\theequation}{A.\arabic{equation}}
\section{Effects of optical lattice potential}\label{appa}

In this appendix, as a subsidiary work, we study a continuum GPE describing
cold atomic gas on a 2D optical lattice potential. 
This model is regarded as the continuum version of the GPE of the BHM. 
Here, we investigate the effect of the lattice potential for quantized vortices 
created by the noise and the magnetic field.
The system corresponds to the previous simulation in Sec.~III, that is, 
the interplay of a small magnetic field and a white noise creates vortices in bulk BEC.

\begin{figure*}[t]
\centering
\includegraphics[width=16cm]{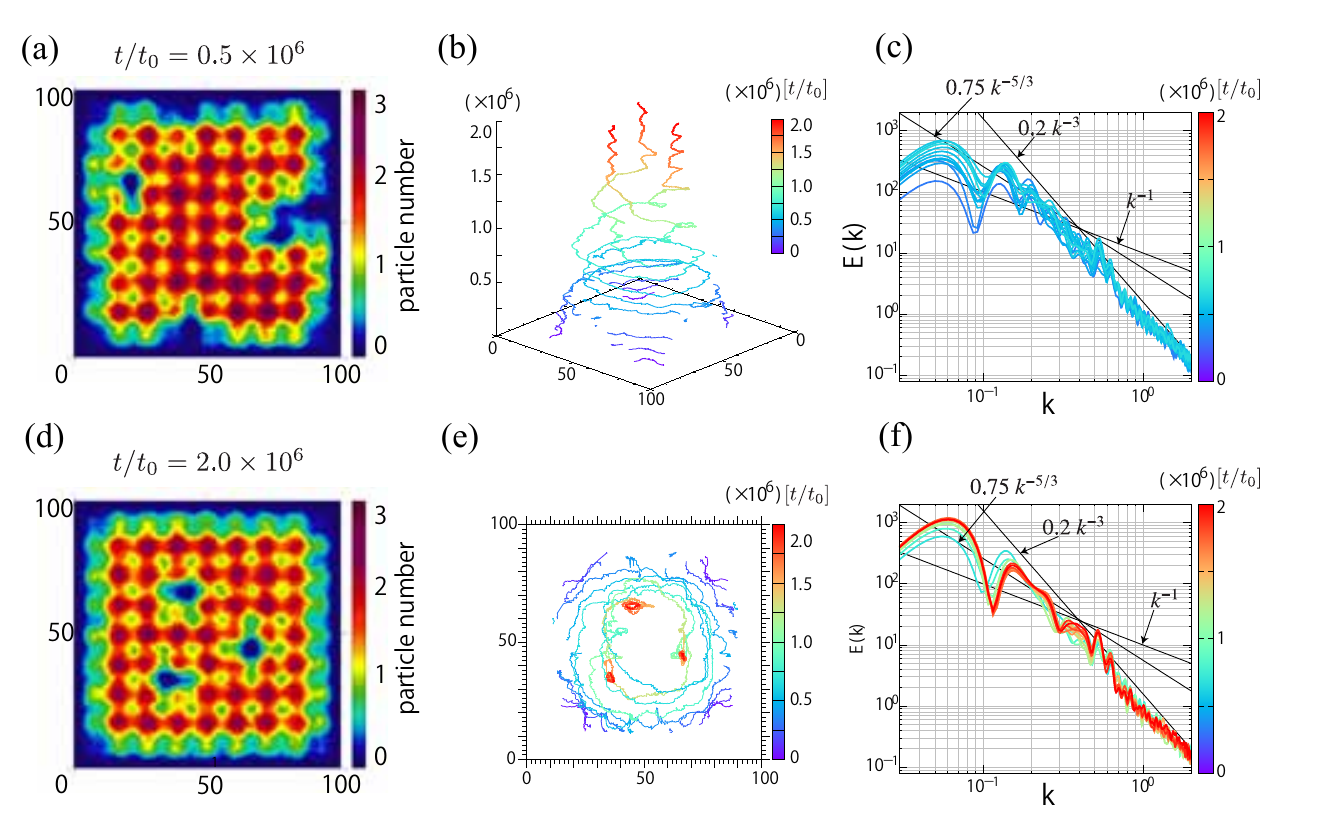}
\caption{(Color online) (a) and (d): Snapshots of density of atom obtained by
solving Eq.(\ref{CGPE2}). 
(b) and (e) Trajectories of intruding vortices.
(c) and (f): the IESs in $0<t/t_{0}<1\times 10^5 $ and $1\times 10^5 <t/t_{0}$, 
where for $0<t/t_{0}<1\times 10^5 $ the system exhibits 
the large cyclotron motion of three vortices, 
whereas for $1\times 10^5 <t/t_{0}$ case large cyclotron motion 
of vortices are suppressed. $t_{0}\sim 7.2\times 10^{-5}$[msec].}
\label{conGP}
\end{figure*}

Let us switch to a continuum model that takes into account effects 
of the optical lattice by a potential. 
The model is obtained by introducing the continuum order parameter field 
$\Phi ({\bf x},t)$.
Then the model is given by the following GPE as in Ref.\cite{Pethick},
\begin{eqnarray}
i\hbar \frac{\partial \Phi ({\bf x},t)}{\partial t}&=&\biggl[-\frac{\hbar^{2}(\nabla-{\bf A}({\bf x},t))^{2} }{2m}+V_{\rm ol}({\bf x}) \nonumber\\
&&\hspace{0.5cm}-\mu+g|\Phi({\bf x},t)|^{2}\biggr]\Phi ({\bf x},t), \label{CGPE} \\
V_{\rm ol}({\bf x})&=&\frac{V_{0}}{2}\biggr[\cos^{2}(kx)+\cos^{2}(ky)\biggr],
\label{OLP}
\end{eqnarray}  
where 
$V_{\rm ol}({\bf x})$ is the square optical lattice potential,
$V_{0}$ is optical lattice depth and $k=\pi/d$ is the wave number of the 
standing-wave laser beam to create the optical lattice ($d$ is a optical lattice spacing,$\sim$ 500[nm]).
The other notations in Eq.(\ref{CGPE}) are standard.
Furthermore, we introduce the magnetic field and the white noise as 
in the previous SGPE of the BHM.
Then, the GPE is given as,
\begin{eqnarray} 
(i-\gamma_{d})\hbar\frac{\partial \Phi ({\bf x},t)}{\partial t}&& \nonumber \\
&&\hspace{-3cm}
=\biggl[-\frac{\hbar^{2}(\nabla-{\bf A}({\bf x},t))^{2}}{2m}+V_{\rm ol}({\bf x})-\mu
\nonumber\\
&&\hspace{-2cm}
+\Xi ({\bf x},t)+g|\Phi({\bf x},t)|^{2}\biggr]\Phi ({\bf x},t),\label{CGPE2}\\
\langle \Xi^{*}({\bf x},t)\Xi({\bf x},t')\rangle &=& 2 
\sigma^{2}\delta (t-t')\delta({\bf x}-{\bf x}'),
\label{noisec}
\end{eqnarray}
where $\Xi({\bf x},t)$ is the noise field and the artificial magnetic field is given 
by the vector potential ${\bf A}({\bf x},t)=(2\pi fy,0)$. 
We treat the parameters $\gamma_d$ and $\sigma^{2}$ 
as phenomenological parameters controlling the dissipation and noise.

In numerical simulation, we put the computational-square mesh with interval $0.1d$.
This interval length is sufficiently small compared with the healing length $\zeta$. 
Thus, we can capture the core structure and the dynamical behavior of vortices. 
In numerical-simulating this model, as one of typical experimental examples, 
we consider a gas of ${}^{87}$Rb atom that has 
the physical parameters, $m=1.44\times 10^{-25}$[kg], $g=1.95\times 10^{-15}$[${\rm J\cdot m^{3}}$], 
and we put the optical lattice 
potential depth $V_{0}$ as $2E_{\rm R}$, where $E_{{\rm R}}$ is the recoil energy of ${}^{87}$Rb.
Moreover, system density is set to be unity (almost one particle) in each potential well of $V_{\rm ol}({\bf x})$. 
The system size is $10d \times 10d$. 

In numerical-simulating the model (\ref{CGPE2}), 
we use the same initial state and boundary condition as in Sec.III.
Please note that also in this continuum SGPE, only the weak magnetic field alone, i.e., 
in the absence of the noise,
cannot create stabilized vortices in the bulk as in the previous case in Sec. III. 

The density snapshots in generating vortices on the sytem are shown in 
Fig.~\ref{conGP} (a), where the vortex core are clearly captured as defects of density. 
After a transient time with cyclotron motions, the vortices are pinned around potential maximum points.
The density snapshot is shown in Fig.~\ref{conGP} (d).
Figures \ref{conGP} (b) and (e) show the real-time dynamics of intruding vortices. 
These trajectories indicate that after a certain amount of time evolution,
the position of the core of the intruding vortex are affected on the optical lattice potential, i.e., vortices are pinned around potential maximum points.
This vortex behavior is similar to the result obtained 
by the previous work of the SGPE of the BHM in Sec. III, 
though the number of intruding vortex is different.

Figures. \ref{conGP} (c) and (f) indicate the IESs. 
Here, the behavior and the appearance of the power law of the IES are quite similar to the previous ones in Figs.~\ref{IES3} (c) and (f).
While vortices are intruding the bulk SF with cyclotron motion, the IES exhibits the $k^{-5/3}$ power low. 
However, once vortices are pinned around potential maximum points of optical lattice 
and forms, i.e., the  Abrikosov lattice, 
the Kolmogorov $k^{-5/3}$ spectrum changes to $k^{-1}$ power law.



\begin{references}
\bibitem{systemsize}
This oscillation is caused by finite system size effect.
\end{references}
\end{document}